\documentclass[twocolumn,aps,prl,showpacs,superscriptaddress]{revtex4-1}
\usepackage[dvipdf]{graphicx}
\usepackage{slashed}
\usepackage{bm}
\begin{document}
\title{ Quantum Spin Hall Effect in Electric and Magnetic Fields without Spin-Orbit Coupling  }
\author{Aiying Zhao}
\email{ayzhao0909@knights.ucf.edu}
\affiliation{University of Science and Technology Beijing,  Beijing 100083, China}
 \affiliation{University of Central Florida,  Orlando, FL 32816-2385 USA}
\author{Qiang Gu}\email{qgu@ustb.edu.cn, corresponding author}\affiliation{University of Science and Technology Beijing, Beijing 100083, China}\author{Richard A. Klemm}
\email{richard.klemm@ucf.edu, corresponding author} \affiliation{University of Central Florida, Orlando, FL 32816-2385 USA}
\date{\today}
\begin{abstract}
From the Dirac equation of an electron in an anisotropic conduction band,
the anisotropy of its motion dramatically affects its interaction with applied electric and magnetic fields.  The  quantum spin Hall  effect (QSHE) is observable in two-dimensional metals without spin-orbit coupling. The dimensionality of the Zeeman interaction plays an important role in the QSHE, and profoundly modifies many interpretations  of measurements of the Knight shift and of  the upper critical field in highly anisotropic superconductors. \end{abstract}
\pacs{05.20.-y, 75.10.Hk, 75.75.+a, 05.45.-a} \vskip0pt\vskip0pt
\maketitle
There has been a very large interest in the QSHE in topological insulators \cite{Kane,Bernevig,Hasan,Qi,Wu}.
In most of these studies, the model Hamiltonian  was proportional to $({\bm p}\times{\bm E})\cdot{\bm\sigma}$ \cite{Bernevig}, where ${\bm p}$ is the momentum of the electron, ${\bm E}=-{\bm\nabla}\Phi-\partial{\bm A}/{\partial t}$ is the electric field, $\Phi$ and ${\bm A}$ are the electrostatic and magnetic vector potentials, and the components of ${\bm \sigma}$ are the Pauli matrices.  Such a Hamiltonian can represent spin-orbit coupling, but omits the magnetic induction ${\bm B}={\bm\nabla}\times{\bm A}$ directly.  In classical physics, the Hall experiment involves both an applied ${\bm E}$ and an applied magnetic field ${\bm H}$.  As shown in the following, the  quantum spin Hall (QSH) Hamiltonian for a two-dimensional (2D) conductor is a generalization of that compact quantum form that includes but does not require spin-orbit coupling.

  The relativistic kinetic energy of an electron in an orthorhombically anisotropic conduction band may be written as
\begin{eqnarray}
T_a&=&\sqrt{mc^2\sum_{i=1}^3\Pi_i^2/m_i+m^2c^4},
\end{eqnarray}
where  $\Pi_i=p_i+eA_i$, $m_i$, $p_i$, and $A_i$ are the effective mass, momentum, and magnetic vector potential in the $i^{\rm th}$
direction, $m$ is the electron rest mass, $-|e|$ is its charge, and $c$ is the vacuum light speed.  Since  $v_i=p_i/m_i<<c$,  $mc^2$ is  the large energy in $T_a$.

In the Supplementary Materials we derive the covariant Dirac equation for a relativistic electron in an orthorhombically anisotropic conduction band,
and demonstrate that it is invariant under all proper and improper Lorentz transformations\cite{SM}.  From the contravariant form of that anisotropic Dirac equation, we used the Foldy-Wouthuysen transformations to eliminate odd powers of the anisotropic momentum operator ${\cal O}_{a}$ to obtain the non-relativistic form of $H_a=T_a+V({\bm r})$ valid to order $1/(mc^2)^{3}$, where $V({\bm r})=-e\Phi({\bm r})$. \cite{SM,Foldy}.
 The most important parts of the Hamiltonian for an electron in a 2D conductor with $m_1=m_2=m_{||}$ are
\begin{eqnarray} H_{2D}&=&H^{T}_{2D}+H^{Z}_{2D}+H^{QSH}_{2D},
\end{eqnarray}
where $H_{2D}^T=({\bm p}_{||}+e{\bm A}_{||})^2/(2m_{||})$ is the kinetic energy, where ${\bm p}_{||}$ and ${\bm A}_{||}$ are the 2D components of ${\bm p}$ and ${\bm A}$,  respectively,
\begin{eqnarray}
H^Z_{2D}&=&\mu_{B||}\sigma_{\perp}B_{\perp},
\end{eqnarray}
is the 2D version of the Zeeman energy, where $\mu_{B||}=e\hbar/(2m_{||})$ is the effective Bohr magneton in 2D, $\hbar=h/(2\pi)$, where $h$ is Planck's constant, $\sigma_{\perp}$ and $B_{\perp}$ are the Pauli matrix and magnetic induction normal to the 2D conductor, and
\begin{eqnarray}
H^{QSH}_{2D}&=&\frac{\mu_{B||}}{2mc^2}[{\bm E}\times({\bm p}+e{\bm A})]_{\perp}\sigma_{\perp},
\end{eqnarray}
is the full version of QSH Hamiltonian in 2D.  There are two parts to this Hamiltonian, the first part proportional to $({\bm E}\times{\bm p})_{\perp}\sigma_{\perp}$, and the second part proportional to $({\bm E}\times e{\bm A})_{\perp}\sigma_{\perp}$.  The focus of previous  QSH work  has been on the first part,
which is the spin-orbit part of the QSH Hamiltonian. When only this term was studied, no magnetic field ${\bm H}$ was applied, so $H^Z_{2D}$ was absent and $H^T_{2D}$ was assumed independent of ${\bm A}_{||}$.

However, the second term in $H^{QSH}_{2D}$ has been overlooked by the community, and it is at least as important.
This term involves both the applied ${\bm E}$ and ${\bm A}$ \cite{AB}, so its inclusion requires the simultaneous inclusions of $H^{T}_{2D}$ and $H^{Z}_{2D}$.  The experimenter has several tools to employ.  Setting a potential difference across the 2D metal leads to  ${\bm E}$ in a fixed direction.  One then applies ${\bm H}$ in the plane normal to the 2D film while containing ${\bm E}$.  One can rotate ${\bm H}$ an angle $\Theta$ from ${\bm E}$ within that plane.  Although it is difficult to control ${\bm A}$, it must  have a component in the 2D film that is normal to ${\bm E}$.  When ${\bm H}$ has a component normal to the 2D film, the Zeeman energy is different for the two electron spin states. But when  ${\bm H}$ is parallel or antiparallel to ${\bm E}$ there is no explicit Zeeman energy, but for ${\bm A}$ having a component normal to ${\bm B}$ and to ${\bm E}$ that is also within the plane, the QSHE can be realized.  Flipping the direction of either ${\bm H}$ or ${\bm E}$ will flip the electron spins, and this can be measured in a number of ways.  One possibility is shown in Fig. 1.
\begin{figure}
\center{\includegraphics[width=0.4\textwidth]{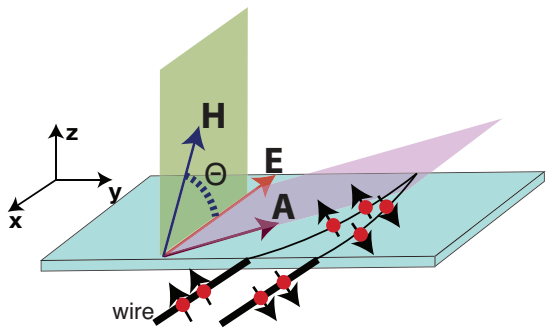}
\caption{Sketch of  a QSHE device in a clean metallic monolayer with applied   ${\bm E}$, ${\bm H}$, and  ${\bm A}$ with a component in the $xy$ plane normal to ${\bm E}$, but no spin-orbit coupling.  From Eq. (4).}}
\end{figure}

In addition, the dimensional dependence of the Zeeman interaction has also been overlooked by many workers in superconductivity. As noted in the Supplementary Materials, in an isotropic 3D metal with $m_1=m_2=m_3=m_g$, $H^Z_{3D}$ is given by
\begin{eqnarray}
H^Z_{3D}&=&\mu_{Bg}{\bm\sigma}\cdot{\bm B},
\end{eqnarray}
where $\mu_{Bg}=e\hbar/(2m_g)$, where $m_g$ is not necessarily equal to $m$ \cite{SM}. In 2D, $H^{Z}_{2D}$ is given by Eq. (3), and  in 1D,
\begin{eqnarray}
H^Z_{1D}&=&0,
\end{eqnarray} since there is no vector product in 1D.  Equations (3), (5), and (6) were overlooked by the superconductivity community, and can explain many  results that were not understood.

Recently, the temperature $T$ dependence of the upper critical magnetic induction $B_{c2,||}(T)$  parallel to atomically thin layers of a variety of gated transition metal dichalcogenide superconductors \cite{Xi,Lu,Fatemi,Sajadi}, of superconducting twisted graphene bilayers \cite{Cao}, and of several organic and heavy fermion superconductors was studied \cite{Agosta,Matsuda}.  In many of these cases, $B_{c2,||}(0)$ was found to greatly exceed the ``Pauli limit'' $B_P=1.86 T_c$ (T/K), where $T_c$ is the superconducting transition temperature in K at ${\bm B}=0$.  That limit assumed that the Zeeman energy splitting $\mu_BB$ between the spin-singlet Cooper pairs exceeded the superconducting gap energy $\Delta(0)$ at $T=0$.  There have been two standard models for this $B_P$ violation.  In the  standard model of layered superconductors \cite{KLB}, the very strong spin-orbit scattering rate $\hbar/\tau_{\rm so}$  was assumed comparable to the total scattering rate $\hbar/\tau$ in the dirty limit, for which the mean-free path $\ell=v_F\tau<<\xi_0$, where $\xi_0$ and $v_F$ are respectively the superconducting coherence length at $T=0$ and the Fermi velocity, both parallel to  the layers.  The second model is that of a thermodynamic phase transition into a  low $T$, high $B(T)$ phase, known as the  Fulde-Farrell-Larkin-Ovchinnikov (FFLO) state \cite{FF,LO}.  This state was predicted to have a gap function $\Delta({\bm r},T)=\Delta(T)e^{i{\bm q}\cdot{\bm r}}$, with a periodic spatial dependence that could only occur in the clean limit $\ell\gg\xi_0$.

In both of those  models, the Zeeman interaction was assumed to be that of a free electron moving isotropically in three spatial dimensions (3D). On a macroscopic scale, the size of an atom is a ``zero-dimensional'' (``0D'') point, as sketched in Fig. 2A. Microscopically, however,  its nucleus  moves slowly inside a 3D electronic shell, and as for the Dirac equation of a free electron, the 3D relativistic motion of each of its neutrons and protons leads to it having an overall  spin $I$ and a nuclear Zeeman energy that can be probed by a time $t$-dependent external magnetic field ${\bm H}(t)$ in nuclear magnetic resonance (NMR) and in Knight shift measurements when in a metal \cite{Hall,Klemm}.  The orbital electrons bound to that nucleus  also move in a nearly isotropic 3D environment,  and have a much  larger Zeeman interaction with ${\bm H}(t)$, modified only by the $V({\bm r})=-e\Phi({\bm r})$  of nearby atoms.

However, when an atomic electron is excited into a crystalline conduction band, it leaves that atomic site and moves with wave vector ${\bm k}$ across the crystal. Its motion depends upon the crystal structure, and can be highly anisotropic. In an isotropic, 3D metal,   $E({\bm k})=\hbar^2{\bm k}^2/(2m)$ for free electrons.  These states are filled at $T=0$ up to the Fermi energy $E_F$ and $H^Z_{3D}=\mu_B{\bm \sigma}\cdot{\bm B}$.  However, in   Si and Ge \cite{Mahan}, the lowest energy conduction bands can be expressed as $E({\bm k})=\hbar^2\sum_{i=1}^3(k_i-k_{i0})^2/(2m_i)$ about some minimal point ${\bm k}_0$, and the $m_i$ can differ significantly from $m$.

In a purely one-dimensional (1D) metal, the conduction electrons move rapidly along the chain of atomic sites, as sketched  in Fig. 2B, usually with a tight-binding 1D band $E(k)$ as sketched in Fig. 2C,  and $H^{Z}_{1D}=0$.   When an electron is in a quasi-1D superconductor such as tetramethyl-tetraselenafulvalene hexafluorophosphate, (TMTSF)$_2$PF$_6$ \cite{Lee}, $E({\bm k})$ is highly anisotropic, the transport normal to the conducting chains is  by weak hopping, so the effective masses in those directions greatly exceed $m$.

 Similarly, in 2D metals, such as monolayer or gated NbSe$_2$, MoS$_2$, WTe$_2$ \cite{Xi,Lu,Fatemi,Sajadi}, and twisted bilayer graphene \cite{Cao}, the effective mass normal to the conducting plane is effectively infinite.  As sketched in Fig. 3, the direction of ${\bm H}$ is very important.  When ${\bm H}$ is  normal to that plane, as in Fig. 3A, the spins of the conduction electrons eventually align either parallel or anti-parallel to ${\bm H}$, giving rise to a Zeeman interaction that can differ from that of a free electron only by the effective mass $M_{||}$. There are two energy dispersions $E({\bm k})$ for up and down spin conduction electrons, as sketched in Fig. 3B.

\begin{figure}
\center{\includegraphics[width=0.45\textwidth]{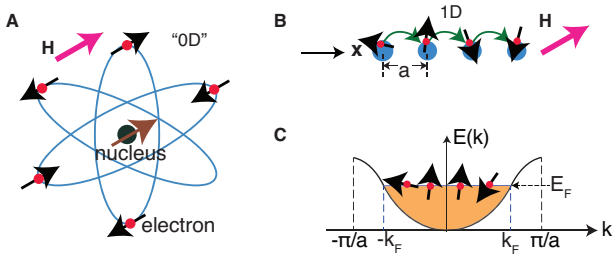}
\caption{{\bf  Atomic (``0D'') components with a full $H^Z_{3D}$ and 1D electronic motion with $H^Z_{1D}=0$.} ({\bf A}) Sketch of an atom of effective point size (``0D''), in which the nuclear components and the electrons move in essentially isotropic 3D environments.  ({\bf B}) An electron moving on a 1D with $H^Z_{1D}=0$. ({\bf C}) The tight-binding model for 1D motion, with the energy levels $E(k)$ filled up to $E_F$.}}
\end{figure}

 However, when ${\bm H}$ lies within the 2D conduction plane, as sketched in Fig. 3C, the Zeeman interactions vanish, so their spin states are effectively random, and there is only one conduction band, as sketched in Fig. 3D.

  In Fig. 3E, sketches of the generic behavior expected for the upper critical induction ${\bm B}_{c2}(T)$ for ${\bm B}=\mu_0{\bm H}$ applied parallel and perpendicular to a 2D film.  The red dashed horizontal line is the effective Pauli limiting induction ${B}^{\rm eff}_P$, which is proportional to the effective mass $m_{||}$ within the conducting plane, and can therefore be either larger or smaller than the result (1.86 $T_c$ T/K) for an isotropic  superconductor.  However, ${\bm B}_{c2,||}(T)$ generically follows the Tinkham thin film formula ${\bm B}_{c2,||}(T)=\mu_0\sqrt{3}\Phi_0/[\pi s\xi_{||}(T)]$ \cite{Klemmbook}, where $s$ is the film thickness, $\Phi_0=h/(2e)$ is the superconducting flux quantum and $\xi_{||}(T)$ is the Ginzburg-Landau coherence length parallel to the film.  There is no Pauli limiting for this ${\bm H}$ direction, consistent with many  experiments \cite{Xi,Lu,Fatemi,Sajadi,Cao,Klemmbook}.

 The Knight shift is the relative change in the NMR frequency for a nuclear species when it is in a metal (or superconductor) from when it is in an insulator or vacuum.  In both cases, the nuclear spin of an atom interacts with that of  one of its orbital electrons via the hyperfine interaction.  But when that atom is in a metal, the orbital electron can sometimes be excited into the conduction band, travelling throughout the crystal, and then returning to the same nuclear site, producing the leading order contribution to the Knight shift \cite{Hall,Klemm}.  The dimensionality of the motion of the electron in the conduction band is therefore crucial in interpreting Knight shift measurements of anisotropic materials, as first noticed in the anisotropic three-dimensional superconductor, YBa$_2$Cu$_3$O$_{7-\delta}$ \cite{Barrett}.

 \begin{figure}
\center{\includegraphics[width=0.45\textwidth]{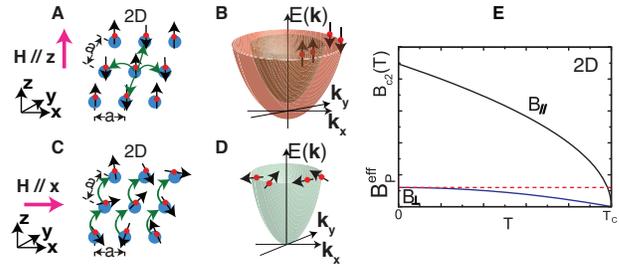}
\caption{{\bf 2D motion, its anisotropic $H^Z_{2D}$, and anisotropic upper critical field} ({\bf A}) Sketch of a 2D ionic lattice in the $xy$ plane with ${\bm H}||{\bm z}$.  The electron  spins experience a full $H^{Z}_{2D,\perp}$,  ({\bf B}) Sketches of $E({\bm k})$ for both spins parallel and antiparallel to ${\bm H}||{\bm z}$. ({\bf C}) Sketch of the same 2D ionic lattice with ${\bm H}||{\bm x}$. The electron form spins have $H^{Z}_{2D,||}=0$. ({\bf D})  Sketch of the single $E({\bm k})$ for both spin states with ${\bm H}||{\bm x}$.
({\bf E}) Sketch of ${\bm B}_{c2,||}(T)$ (black), ${\bm B}_{c2,\perp}(T)$ (blue), and the renormalized Pauli limit $B_{P}^{\rm eff}$ (red dashed)\cite{SM}.}}
\end{figure}

 In Knight shift $K(T)$ measurements with  ${\bm H}$ applied parallel to the layers of  Sr$_2$RuO$_4$, $H^{Z}_{2D}$ should be vanishingly small, so one expects little change in $K(T)$ at and below $T_c$, due to Eq. (3), as observed \cite{Ishida}. Similarly, Eq. (6) implies that $K(T)$ on the quasi-one-dimensional superconductor (TMTSF)$_2$PF$_6$ should be nearly constant, as observed \cite{Lee}.    A recent $K(T)$ measurement on Sr$_2$RuO$_4$ under uniaxial planar pressure did show a substantial $K(T)$ variation below $T_c$ \cite{Pustogow},  in agreement with scanning tunneling microscopy results of a nodeless superconducting gap \cite{Suderow}.

 In conclusion, a new quantum spin Hall effect is predicted for monolayer metallic films that makes use of ${\bm E}$ and ${\bm A}$ but not of spin-orbit coupling. A simple explanation for the strong violation of the ``Pauli limit'' in experimental measurements of $B_{c2,||}(T)$ in clean ultrathin superconductors is provided.   Knight shift measurements of highly anisotropic superconductors should not be interpreted as if they were isotropic.


\section*{Acknowledgments}
 This work was supported by the National Natural Science Foundation of China through Grant no. 11874083.  AZ was also supported by the China Scholarship Council.  We thank Luca Argenti, Madhab Neupane, and Jingchuan Zhang for helpful  discussions.
 
 \section{Appendix}
 A relativistic electron in an anisotropic environment of orthorhombic symmetry satisfies the Schr{\"o}dinger equation based upon the modified Hamiltonian $H_a=T_a+V$, where $T_a$ is given by Eq. (1) in the text and $V({\bm r})=-e\Phi({\bm r})$ \cite{Dirac,BjorkenDrell},
\begin{eqnarray}
i\hbar\frac{\partial\psi}{\partial t}&=&\left[c{\bm\alpha}_a\cdot{\bm\Pi}+\beta mc^2-e\Phi\right]\psi=H_a\psi,
\end{eqnarray}
where ${\bm\Pi}={\bm p}+e{\bm A}$, ${\bm p}\rightarrow-i\hbar{\bm\nabla}$, and
\begin{eqnarray}
\alpha_a^{\mu}&=&\left[\begin{array}{cc}0&\sigma_a^{\mu}\\
\sigma_a^{\mu}&0\end{array}\right],\sigma_a^{\mu}=\frac{\sigma^{\mu}}{\sqrt{m_{\mu}/m}},\beta=\left[\begin{array}{cc}1&0\\0&-1\end{array}\right],
\end{eqnarray}
for $\mu=1,2,3$, the $\sigma^{\mu}$ are the Pauli matrices, and both the $\alpha^{\mu}_a$ and $\beta$ are rank-4 matrices, where 1 represents the rank-2 identity matrix.
  We used subscripts for the contravariant forms of the effective masses, in order to avoid confusion with superscripts representing exponents, which appear subsequently.  The matrices satisfy $\left\{\alpha_a^{\mu},\alpha_a^{\nu}\right\}=2\delta^{\mu\nu}m/m_{\mu}$, and $\left\{\alpha_a^{\mu},\beta\right\}=0$.

  From Eq. (1), the $\mu^{\rm th}$ component of the probability  current  is   $j^{\mu}=\psi^{\dagger}\alpha_a^{\mu}\psi$, and since $\rho=\psi^{\dag}\psi$, the continuity equation
$\frac{\partial\rho}{\partial t}+\frac{\partial}{\partial x^{\mu}}j^{\mu}=\frac{\partial\rho}{\partial t}+{\rm div}{\bm j}=0$,
 is still satisfied with effective mass anisotropy.

\section{Covariant anisotropic Dirac equation}
 To demonstrate the Lorentz invariance of this anisotropic Dirac equation, we multiply its contravariant form, Eq. (1), by $\beta$,
\begin{eqnarray}
\gamma^{0}_a&=&\beta=\left[\begin{array}{cc}1&0\\0&-1\end{array}\right],\hskip15pt\gamma_a^{\mu}=\beta\alpha_a^{\mu}=\left[\begin{array}{cc}0&\sigma_a^{\mu}\\ -\sigma_a^{\mu}&0\end{array}\right],\>\>\>\>
\end{eqnarray}
for $\mu=1,2,3$.  We note that $\gamma_a^0$ is hermitian, so that $(\gamma_a^0)^2=1\equiv g_a^{00}$.  The $\gamma_a^{\mu}$ for $\mu=1,2,3$ satisfy the anticommutation relations
\begin{eqnarray}
\left\{\gamma_a^{\mu},\gamma_a^{\nu}\right\}&\equiv&2g_a^{\mu\nu}\delta^{\mu\nu}=\frac{-2\delta^{\mu\nu}m}{m_{\mu}}.
\end{eqnarray}
 These features lead to the metric $g_a$ given by
\begin{eqnarray}
g_a&=&\left(\begin{array}{cccc}1&0&0&0\\
0&-m/m_1&0&0\\
0&0&-m/m_2&0\\
0&0&0&-m/m_3\end{array}\right).
\end{eqnarray}
We then may use the Feynman slash notation \cite{BjorkenDrell},
\begin{eqnarray}
\slashed{\nabla}_a&=&\gamma_a^{\mu}\frac{\partial}{\partial x^{\mu}}=\frac{\gamma_a^0}{c}\frac{\partial}{\partial t}+{\bm\gamma}_a\cdot{\bm\nabla},\nonumber\\
\slashed{A}_a&=&\gamma_a^{\mu}A_{\mu}=\gamma_a^0A^0-{\bm\gamma}_a\cdot{\bm A},
\end{eqnarray}
to write the anisotropic Dirac equation in covariant form,
\begin{eqnarray}
(i\hbar \slashed{\nabla}_a+e \slashed{A}_a-mc)\psi&=&0.
\end{eqnarray}

  Here we demonstrate that the norm for a relativistic electron in an orthorhombically anisotropic conduction band with metric $g_a$ given by Eq. (5) is indeed invariant under the most general proper Lorentz transformation $a$, find the matrix form of $a$, and show that it  has O(3,1) symmetry, as for the isotropic case. Examples of  general rotations and general boosts are given. Improper Lorentz transformations such as reflections, parity, charge conjugation,  and time inversion can be treated exactly as for the isotropic Dirac equation, as shown in the following \cite{BjorkenDrell,Jackson}.
\section{Proof of covariance}
For a general proper Lorentz transformation in a relativistic orthorhombic system,
$x'=ax$, where $x'$ and $x$ are column  (Nambu) four-vectors and $a$ is the appropriate proper anisotropic Lorentz transformation, which is to be found based upon symmetry arguments.  We require the norm with $g_a$ to be invariant under all possible Lorentz transformations \cite{Jackson}
$(x,g_ax)=(x',g_ax')$,
or
$x^{T}g_ax=(x')^{T}g_ax'$,
where $x^T$ is the transpose (row) form of the four-vector $x$ and  $g_a$ is given by Eq. (8).
We then have
$x^Tg_ax=(x')^Tg_ax'=x^Ta^Tg_aax$,
 which implies
$g_a=a^Tg_aa$.
As for the isotropic case, we assume
$a=e^{L_a}$,
so that
$a^T=e^{L_a^T}$, and
$a^{-1}=e^{-L_a}$.
Then from $g_a=a^Tg_aa$, we have $g_aa^{-1}=a^Tg_a$ and hence that $a^{-1}=g_a^{-1}a^Tg_a$.  We then may rewrite this as
\begin{eqnarray}
e^{-L_a}&=&g_a^{-1}e^{L_a^T}g_a=e^{g_a^{-1}L_a^Tg_a}.
\end{eqnarray}
 Taking the logarithm of both sides, we obtain
$-L_a=g_a^{-1}L_a^Tg_a$, or that
$-g_aL_a=L_a^Tg_a=(g_aL_a)^T$,
which requires  $g_aL_a$ to be  antisymmetric.  We then write \cite{Jackson}
\begin{eqnarray}
L_a&=&\left(\begin{array}{cccc}0&\frac{-\zeta_1\sqrt{m}}{\sqrt{m_1}}&\frac{-\zeta_2\sqrt{m}}{\sqrt{m_2}}&\frac{-\zeta_3\sqrt{m}}{\sqrt{m_3}}\\
&&&\\
\frac{-\zeta_1\sqrt{m_1}}{\sqrt{m}}&0&\frac{\omega_3\sqrt{m_1}}{\sqrt{m_2}}&\frac{-\omega_2\sqrt{m_1}}{\sqrt{m_3}}\\
&&&\\
\frac{-\zeta_2\sqrt{m_2}}{\sqrt{m}}&\frac{-\omega_3\sqrt{m_2}}{\sqrt{m_1}}&0&\frac{\omega_1\sqrt{m_2}}{\sqrt{m_3}}\\
&&&\\
\frac{-\zeta_3\sqrt{m_3}}{\sqrt{m}}&\frac{\omega_2\sqrt{m_3}}{\sqrt{m_1}}&\frac{-\omega_1\sqrt{m_3}}{\sqrt{m_2}}&0\end{array}\right),\nonumber\\
\end{eqnarray}
for which $g_aL_a$ is easily shown to be antisymmetric, and $L_a$ may be written as

\begin{eqnarray}
L_a&=&-{\bm\omega}\cdot{\bm S}_a-{\bm\zeta}\cdot{\bm K}_a,
\end{eqnarray}
where  each component four-vector of ${\bm S}_a$ and of ${\bm K}_a$ has only two non-vanishing elements.
It is easy to show that
$\left[S_{ai},S_{aj}\right]=\epsilon_{ijk}S_{ak}$,
$\left[K_{ai},K_{aj}\right]=-\epsilon_{ijk}S_{ak}$, and
$\left[S_{ai},K_{aj}\right]=\epsilon_{ijk}K_{ak}$,
so the anisotropic Lorentz transformation matrix $L_a$ has SL(2,C) or O(3,1) group symmetry, precisely as for the isotropic case \cite{Jackson}.

We now provide some  examples.  We define
$\omega=\sqrt{\omega_1^2+\omega_2^2+\omega_3^2}$ and
\begin{eqnarray}
A_i&=&\cos\omega+\frac{\omega_i^2}{\omega^2}(1-\cos\omega),\\
B_{ijk}^{\pm}&=&\Bigl(\frac{m_i}{m_j}\Bigr)^{1/2}\Bigl[\frac{\omega_i\omega_j}{\omega^2}(1-\cos\omega)\pm\frac{\omega_k}{\omega}\sin\omega\Bigr].
\end{eqnarray}
Then, for a general rotation,
\begin{eqnarray}
e^{-{\bm\omega}\cdot{\bm S}_a}&=&
\left(\begin{array}{cccc}1&0&0&0\\
0&A_1&B^{+}_{123}&B^{-}_{132}\\
0&B_{213}^{-}&A_2&B^{+}_{231}\\
0&B^{+}_{312}&B^{-}_{321}&A_3\end{array}\right),
\end{eqnarray}
the determinant of which is 1, as required for a general rotation.

For the general boost case, we first set ${\bm \zeta}=({\bm \beta}/{\beta})\tanh^{-1}\beta$, where ${\bm\beta}={\bm v}/c$, (unrelated to the matrix $\beta$ in Eq. (2)) ${\bm v}$ is the electron's velocity, and define
$\zeta=\sqrt{\zeta_1^2+\zeta_2^2+\zeta_3^2}$,
$\cosh\zeta=\gamma=\frac{1}{\sqrt{1-\beta^2}}$,
$\sinh\zeta=\gamma\beta$, and
$\beta=\sqrt{\beta_1^2+\beta_2^2+\beta_3^2}$, as for the isotropic case \cite{Jackson}.  Then we define
\begin{eqnarray}
C_{i}^{\pm}&=&-\gamma\beta_i(m_i/m)^{\pm 1/2},\\
D_{i}&=&1+\frac{(\gamma-1)\beta_i^2}{\beta^2},\\
E_{ij}&=&(\gamma-1)\frac{\beta_i\beta_j}{\beta^2}\Bigl(\frac{m_i}{m_j}\Bigr)^{1/2}.
\end{eqnarray}
Then for the general boost case, we have \cite{Jackson}
\begin{eqnarray}
e^{-{\bm\zeta}\cdot{\bm K}_a}
&=&\left(\begin{array}{cccc}\gamma&C_1^{-}&C_2^{-}&C_3^{-}\\
C_1^{+}&D_1&E_{12}&E_{13}\\
C_2^{+}&E_{21}&D_2&E_{23}\\
C_3^{+}&E_{31}&E_{32}&D_3\end{array}\right),
\end{eqnarray}
the determinant of which is  also 1,
as required for a general boost.

Hence  Eq. (7) is invariant under the most general proper Lorentz transformation.  As argued in the following, it is also invariant under all of the relevant improper Lorentz transformations:  reflections or parity, charge conjugation, and time reversal \cite{BjorkenDrell}.

Before we demonstrate the proof of invariance of the anisotropic Dirac equation under all possible improper Lorentz transformations, it is useful to employ the Klemm-Clem transformations that were used to transform such an anisotropic Ginzburg-Landau model into isotropic form \cite{KlemmClem}.  To do so, we first make the anisotropic scale transformation of the spatial parts of the contravariant form of the anisotropic Dirac equation, Eq. (1),
\begin{eqnarray}
\frac{\partial}{\partial x^{\mu}}&=&\sqrt{m_{\mu}/m}\frac{\partial}{\partial x^{\mu'}},\\
A^{\mu}&=&\sqrt{m_{\mu}/m}A^{\mu'},
\end{eqnarray}
which transforms Eq. (1) to
\begin{eqnarray}
i\hbar\frac{\partial\psi}{\partial t}&=&\left[c{\bm\alpha}\cdot{\Pi}'+\beta mc^2-e\Phi\right]\psi=H\psi,
\end{eqnarray}
where
\begin{eqnarray}
\alpha^{\mu}&=&\left[\begin{array}{cc}0&\sigma^{\mu}\\ \sigma^{\mu}&0\end{array}\right],
\end{eqnarray}
which is precisely the same form as the isotropic Dirac equation.

It is easy to show  that this transformation preserves the Maxwell equation of no monopoles, ${\bm\nabla}'\cdot{\bm B}'=0$, provided that
\begin{eqnarray}
B^{\mu}&=&\sqrt{m/m_{\mu}}B^{\mu'},
\end{eqnarray}
which is easy to show preserves the required relation
\begin{eqnarray}
{\bm B}'&=&{\bm\nabla}'\times{\bm A}'.
\end{eqnarray}
We note that ${\bm B}'$ is no longer parallel to ${\bm B}$, but can be made parallel to it by a proper rotation\cite{KlemmClem}.  The magnitude $|{\bm B}'|$ can then be made equal to $|{\bm B}|$ by an isotropic scale transformation \cite{KlemmClem}.

Hence, it is then easy to construct the transformed covariant form of the anisotropic Dirac equation, as it has exactly the same form as does the isotropic covariant form of the Dirac equation.  Hence, the proof of covariance under the three types of improper Lorentz transformations, reflections, charge conjugation, and time reversal, follow by inspection.

\section{Expansion about the Non-relativistic Limit}
We used   the Foldy-Wouthuysen transformations to eliminate the odd terms in the anisotropic operator ${\cal O}={\bm\alpha}_a\cdot{\bm\Pi}$ obtained in the power series in $(mc^2)^{-1}$  to obtain the non-relativistic limit of $H_a$ in Eq. (1) \cite{BjorkenDrell,Foldy}.  To order $(mc^2)^{-3}$, ${H}^{NR}= \beta mc^2+\delta{H}^{NR}$, where
\begin{eqnarray}
&&\delta{H}^{NR}=\beta\Biggl\{\sum_{\mu=1}^3\biggl(\frac{\Pi_{\mu}^2}{2m_{\mu}}
+\frac{\mu_Bm\sqrt{m_{\mu}}\sigma_{\mu}B_{\mu}}{(m_g)^{3/2}}\biggr)  \nonumber  \\
&&-\frac{1}{2mc^2}\biggl[\sum_{\mu=1}^3\biggl(\frac{\Pi_{\mu}^2}{2m_{\mu}}
+\frac{\mu_Bm\sqrt{m_{\mu}}\sigma_{\mu}B_{\mu}}{(m_g)^{3/2}}\biggr)\biggr]^2
 +\frac{\mu_B^2}{2c^4}\sum_{\mu=1}^3\frac{E_{\mu}^2}{m_{\mu}}\Biggr\}  \nonumber\\
&&+\frac{\mu_B}{4c^2}\sum_{\mu=1}^3\frac{\hbar}{m_{\mu}}\frac{\partial E_{\mu}}{\partial x_{\mu}} +\frac{\mu_B}{4c^2(m_g)^{3/2}}\sum_{\mu,\nu,\lambda=1}^3
\Bigl[\Bigl(2E_{\mu}\Pi_{\nu}+i\hbar\frac{\partial E_{\mu}}{\partial x_{\nu}}\Bigr)\nonumber\\
&&
\epsilon_{\mu\nu\lambda}\sqrt{m_{\lambda}}\sigma_{\lambda}\Bigr]-e\Phi,\label{HNR}
\end{eqnarray}
where $\Pi_i=p_i+eA_i$, the geometric mean $m_g=(m_1 m_2 m_3)^{1/3}$,  and $\epsilon_{\mu\nu\lambda}$ is the Levi-Civita symbol. The Hamiltonian for an electron or hole is obtained respectively with $\beta=1,-1$ \cite{BjorkenDrell}.  Due to its importance for the spin Hall effect, we remark that in the penultimate term in $\delta{H}^{NR}$, Foldy and Wouthuysen included the $E_{\mu}\Pi_{\nu}$ term but omitted the $i\hbar(\partial E_{\mu}/\partial x_{\nu})$ term\cite{Foldy}. Subsequently, Bjorken and Drell included both terms, but omitted the $A_{\nu}$ part of $\Pi_{\nu}$\cite{BjorkenDrell}.  We emphasize that ${\bm A}$ and $\Phi$ are the important quantum mechanical potentials, as ${\bm B}$ can vanish in regions where ${\bm A}\ne0$, as noted in the text. This term leads to the quantum spin Hall effect in a 2D conductor, given by Eq. (4) in the text.

Here we describe the most important effects of motion dimensionality. We define $\delta H_{a}\equiv H_{a}-mc^2$.  An electron in an isotropic 3D conduction band with $m_1=m_2=m_3=m_g$ satisfies the Schr{\"o}dinger equation $H^{NR}_{a,3D}\psi=i\hbar(\partial\psi/\partial t)$, where
\begin{eqnarray}
&&\delta H^{NR}_{a,3D}=\frac{{\bm \Pi}^2}{2m_g}+\mu_{Bg}{\bm \sigma}\cdot{\bm B}-\frac{1}{2mc^2}\Bigl(\frac{{\bm \Pi}^2}{2m_g}+\mu_{Bg}{\bm \sigma}\cdot{\bm B}\Bigr)^2  \nonumber \\
&&+\frac{\mu_B^2}{2m_gc^4}{\bm E}^2
+\frac{\mu_{Bg}}{4mc^2}\Bigl[\hbar{\bm \nabla}\cdot{\bm E}+\Bigl(2{\bm E}\times{\bm\Pi}-i\hbar{\bm\nabla}\times{\bm E}\Bigr)\cdot{\bm\sigma}\Bigr] \nonumber \\
&&-e\Phi({\bm r}),
\end{eqnarray}
where ${\bm E}=-{\bm\nabla}\Phi-\frac{\partial {\bm A}}{\partial t}$  and ${\bm B}={\bm\nabla}\times{\bm A}$ are  the semi-classical electric field and magnetic induction, ${\bm\Pi}=-i\hbar{\bm\nabla}+e{\bm A}$, and $\mu_{Bg}=e\hbar/(2m_g)$ is the 3D effective Bohr magneton.  The Zeeman energy is $\mu_{Bg}{\bm \sigma}\cdot{\bm B}$.

 For  a 2D conductor with $m_1=m_2=m_{||}$ and $m_3\rightarrow\infty$,
\begin{eqnarray}
&& \delta H^{NR}_{a,2D}=\frac{{\bm \Pi_{||}}^2}{2m_{||}}+\mu_{B||}\sigma_{\perp}B_{\perp}-\frac{1}{2mc^2}\Bigl(\frac{\bm \Pi_{||}^2}{2m_{||}} \
+\mu_{B||}\sigma_{\perp}B_{\perp}\Bigr)^2 \nonumber \\
&&+\frac{\mu_{B}^2}{2m_{||}c^4}{\bm E}_{||}^2
+\frac{\mu_{B||}}{4mc^2}\Bigl[\hbar{\bm \nabla}_{||}\cdot{\bm E}_{||}+\Bigl(2{\bm E}\times{\bm\Pi}-i\hbar{\bm\nabla}\times{\bm E}\Bigr)_{\perp}\sigma_{\perp}\Bigr] \nonumber \\
&&-e\Phi({\bm r}_{||}),
 \end{eqnarray}
 where the subscripts $||$ and $\perp$ respectively denote the  components parallel and perpendicular to the film, and $\mu_{B||}=e\hbar/(2m_{||})$ is the effective Bohr magneton.  In a 2D film, a vector product or curl can only exist with both vectors in the film.  The Zeeman interaction, $\mu_{B||}\sigma_{\perp}B_{\perp}$,  only exists for ${\bm B}$ normal to the metallic film.

For an electron in a one-dimensional conduction band with $m_1=m_{||}$, $m_2=m_3\rightarrow\infty$,
\begin{eqnarray}
 \delta H^{NR}_{a,1D}&=&\frac{\Pi^2_{||}}{2m_{||}}-\frac{\Pi_{||}^4}{8mm_{||}^2c^2}+\frac{\mu_B^2}{2m_{||}c^4}E_{||}^2 \nonumber\\
 && +\frac{\hbar\mu_{B||}}{4mc^2}\frac{\partial E_{||}}{\partial r_{||}}-e\Phi(r_{||}),
 \end{eqnarray}
 where the subscript $||$  denotes  the conduction direction,  and $\mu_{B||}=e\hbar/(2m_{||})$.  There is no vector product, and no Zeeman interaction, even to order $(mc^2)^{-4}$.

After this work was completed, we found that most of $\delta{H}^{NR}$ was previously obtained \cite{Safonov}. However, that author did not include part of the second and the third term, each of order $(mc^2)^{-3}$, and omitted the $A_{\nu}$ part in the penultimate term, overlooking the quantum spin Hall effect in a 2D conductor.

We finally remark that we have treated the electromagnetic fields semiclassically, with ${\bm B}={\bm \nabla}\times{\bm A}$ and ${\bm E}=-{\bm \nabla}\Phi-\partial{\bm A}/\partial t$, as was done for isotropic systems \cite{BjorkenDrell,Foldy}. This omits the corrections arising from quantum electrodynamics, whereby an electron can emit and absorb photons, resulting in the anomalous magnetic moment of the electron.

\end{document}